\documentclass[reprint,amsmath,amssymb,aps
]{revtex4-2}
\usepackage{ulem}
\usepackage{graphicx}
\usepackage{dcolumn}
\usepackage{bm}
\makeatletter
\newcommand*{\rom}[1]{\expandafter\@slowromancap\romannumeral #1@}
\DeclareMathOperator{\sech}{sech}

\begin{document}
	
	\preprint{APS/123-QED}
	
	\title{Atomic-Scale Spin-Wave Polarizer Based on a Sharp Antiferromagnetic Domain Wall}
	
\author{Ehsan Faridi$^1$}
\email{efbmm@mail.missouri.edu} 
\author{Se Kwon Kim$^2$}
\author{Giovanni Vignale$^1$}
\email{vignaleg@missouri.com} 

\affiliation{$^1$Department of Physics and Astronomy, University of Missouri, Columbia, Missouri 65211, USA\\ $^2$Department of Physics, Korea Advanced Institute of Science and Technology, Daejeon 34141, Korea}

	\date{\today}

	\begin{abstract}
		
		We theoretically study the scattering of spin waves from a sharp domain wall in an antiferromagnetic spin chain. While the continuum model for an antiferromagnetic material yields the well-known result that spin waves can pass through a wide domain wall with no reflection, here we show that, based on the discrete spin Hamiltonian, spin waves are generally reflected by a domain wall with a reflection coefficient that increases as the domain-wall width decreases. Remarkably, we find that, in the interesting case of an atomically sharp domain wall, the reflection of spin waves exhibits strong dependence on the state of circular polarization of the spin waves, leading to mainly reflection for one polarization while permitting  partial transmission for the other, thus realizing an atomic-scale spin-wave polarizer. The polarization of the transmitted spin wave depends on the orientation of the spin in the sharp domain wall, which can be controlled by an external field or spin torque. Our utilization of a sharp antiferromagnetic domain wall as an atomic-scale spin-wave polarizer leads us to envision that  ultra-small magnetic solitons such as  domain walls and  skyrmions may enable realizations of atomic-scale spin-wave scatterers with useful functionalities.
		
	\end{abstract}
	
	\maketitle

	\section{\label{sec:level1}Introduction}

		The propagation of spin waves in one dimensional magnets is of great fundamental interest and is also relevant for the transport of information in magnetic nanostructures  \cite{ jungwirth2016antiferromagnetic,baltz2018antiferromagnetic}. Compared to conventional electronic spin currents, spin waves in magnetic insulators hold the promise to reduce the energy dissipation owing to the absence of  Joule heating.  In addition, spin waves  in  antiferromagnetic materials naturally occur at higher frequency than  ferromagnetic spin waves, raising hope for fast spintronic applications \cite{ hortensius2021coherent}. In contrast to ferromagnetic materials, where spin waves have only one state of circular polarization  \cite{keeffer1953spin},  antiferromagnetic materials host all types of polarization ranging from circular to elliptical depending on  hard-axis anistropy.\cite{rezende2019introduction,lebrun2020long,kamra2020antiferromagnetic}  
	
	The interaction between  spin waves and spin textures such as domain walls (DW) has been studied extensively in the continuum micromagnetic approximation, where only wide domain walls, much larger than the atomic scale, are considered. It has been shown that, in the absence of a magnetic field, spin waves experience negligible reflection over wide range of frequency when propagating on a static smooth domain wall \cite{kim2014propulsion}. Applying a magnetic field gives rise to a finite magnetization inside the DW which results to a field-controlled reflection of spin waves from a DW \cite{shen2020driving}.  In many studies the interaction between the spin wave and a DW is investigated as a new method to control the transmission of the spin wave 
	\cite{hamalainen2018control,woo2017magnetic,voto2017pinned,fukami2013depinning}.

	The interaction between spin waves and a DW can be substantially different from that obtained in the continuum model as the DW  becomes narrower. When the DW width is comparable to the lattice constant, the continuum approximation breaks down and thus the results obtained with the assumption of smooth textures can be invalidated. For example, in a ferromagnetic spin chain, it has been shown that spin waves experience strong refection from a narrow DW even in the absence of an external field \cite{yan2012magnonic}. However, an analogous investigations has not been conducted for an antiferromagnetic narrow DW. 
    
	Recently, atomically sharp DWs have been observed by electron microscopy  in antiferromagnet CuMnAs  and their existence is found to be consistent with DFT calculations \cite{krizek2020atomically}. These findings   motivate us to pursue the study of  the interaction between spin waves and antiferromagnetic DW within a discrete spin chain model, which allows us to treat atomically sharp DWs. 
	
	The fundamental difference between smooth and sharp AF domain walls is that the latter can be viewed as a point-like ferromagnetic insertion in an otherwise regular antiferromagnetic spin structure. (Similarly, a sharp ferromagnetic domain wall can be viewed as a point-like antiferromagnetic insertion in a regular ferromagnetic structure).

	In this paper we find that, as the antiferromagnetic DW becomes atomically sharp,   left-handed (LH) and right-handed (RH) spin waves get reflected with different amplitudes, depending on the orientation of the spins at the domain boundaries.  Therefore such a sharp DW can act as a {\it polarizer} for spin waves, allowing spin waves of a given polarization to be transmitted much more efficiently than spin waves of the opposite polarization.  Furthermore, the selectivity of the polarizer can be reversed by reversing the orientation of the spins at the center of the domain wall. Lastly, we show that the selectivity of the spin polarizer is a function of the magnetic anisotropy field, and tends to vanish when the latter increases.  
	
	This paper is organized as follows: The model and the theoretical formulation of the problem are outlined in Sec. \rom{2}. In Sec. \rom{3} we present the numerical results for the transmission$/$reflection probabilities of AF spin waves from a DW with an easy axis anisotropy.  The transmission coefficients are shown to depend on the spin-wave wave vector and on its circular polarization.  Here we also discuss the impact of changing magnetic anisotropy on the transmission of  spin waves through the DW.   Technical details are presented in the Appendixes.

	\section{\label{sec:level1}THEORETICAL FORMULATION}
	
	\subsection{\label{sec:level2} Spin wave Hamiltonian}
	
	Our starting point is a quasi-1D antiferromagnetic nanowire with a lattice spacing $a$ in which a DW separates  two homogeneous antiferromagnetic domains on the y-axis. A sketch of a N\'eel type antiferromagnetic DW with two arrows indicating two oppositely oriented  spins on each sub-lattice  is shown in Fig. 1(a). We consider an atomistic Heisenberg Hamiltonian of the form	
	\begin{equation}
		H= J\sum_{n}\textbf{S}_ n \textbf{.} \textbf{S}_ {n+1}-\textit{D}\sum_{n}({S_{  z}^n})^2 \,,
	\end{equation}
where  $S_ n$ is the  spin on the site $n$. The first term describes the isotropic  exchange interaction between neighboring spins:  for $J>0$ this favors antiparallel alignment of neighboring spins. The second term, with $D>0$, describes a uniaxial magnetic anisotropy, which favors the alignment of the spins along the $z$-axis .
	
The equilibrium configuration of a DW between two uniform antiferromagnetic regions can be obtained by minimizing the energy with respect to a set of angles $\theta_n$ describing the equilibrium orientation of $\textbf{S}_n$  relative to the $z$-axis in the $(z,y)$ plane (we assume that the DW lies entirely in this plane).   Requiring the energy to be stationary with respect to infinitesimal variations of $\theta_n$  yields the equations
	\begin{equation}\label{DW_profile}
		\sin(\theta_n-\theta_{n-1})+\sin(\theta_n-\theta_{n+1})-\dfrac{D}{J} \sin(2\theta_n)=0\,.
	\end{equation}
To solve Eq.~(\ref{DW_profile}) we imposed  the boundary condition  that the spins on opposite sides of the DW region are  in the $z$ direction (See Fig. 1(a)).
For small values of the anisotropy, i.e $D/J\ll 1$,  Eq.~(\ref{DW_profile}) can be solved analytically using the approximation $| |\theta_{n+1}-\theta_n|-\pi| \ll 1$.  The solution has the form of an  antiferromagnetic Walker type profile.  As the anisotropy  increases the DW starts to shrink and for $D/J= 2/3$ the spins stay close to their anisotropy axis forming an abruptly sharp DW.

In order to construct a linearized equation of motion for spin excitations on top of the DW, it is convenient to write the Hamiltonian in a local coordinate system which is rotated about $x$-axis, in such a way that,  the local $Z$ axis coincides with the local  orientation of $\textbf{S}_n$ at equilbrium.  	
	
The relation between the components of the spin in the local coordinate system ($X,Y,Z$) and in the global coordinate system ($x,y,z$) is
	\begin{equation}
		\begin{pmatrix}
				S_{ nx}  \\S_{ny} \\S_{nz}
		\end{pmatrix}=	
		\begin{pmatrix}
			1&0&0 \\
			0&\cos{\theta_n}& \sin{\theta_n}\\ 0 & -\sin{\theta_n} & \cos{\theta_n} 
		\end{pmatrix}
		\begin{pmatrix}
			S_{ nX} \\S_{nY} \\S_{ nZ}
		\end{pmatrix}
	\end{equation}
	
We  assume that the magnitudes of $S_{ nX}$ and  $S_{nY}$ (i.e., the non-equilibrium components of the spin in the local reference frame)  are small in comparison with the magnitude of the spin on  site $n$:  $|S_{nX}|,|S_{nY}|\ll |S_{nZ}|$. Expanding to second order  in $S_{ nX}$ and  $S_{nY}$ the Hamiltonian takes the form:	
	
	\begin{equation}
		\begin{aligned}
			H= J\sum_{n}S_{nX}S_{n+1X}+ \qquad \qquad \qquad  \qquad  \qquad \qquad  \qquad  \\ \cos(\theta_n-\theta_{n+1})\left(S_{nY}S_{n+1Y}+S_{nZ}S_{n+1Z}\right)+ \qquad  \qquad       \\
			\sin(\theta_n-\theta_{n+1})   \left(S_{nY}S_{n+1Z}-S_{nZ}S_{n+1Y}\right) - \qquad \qquad          \\
			\textit{D} \left(S_{nZ}\cos\theta_n-S_{nY}\sin\theta_n\right)^2 ,
			\qquad \qquad  \qquad  \qquad  \quad  \quad 
		\end{aligned}
	\end{equation}	
	
Next, we perform the transformation 
	\begin{equation}
		\begin{aligned}
			S_{nX}= \frac{  S_{n+} +  S_{n-}}{2}  \qquad    \qquad     \quad    \\
			S_{nY}= \frac{ S_{n+} -  S_{n-}}{2i}    \qquad   \qquad     \quad   \\
			S_{nZ} \approx  S-\frac{S_{n-} S_{n+} + S_{n+}S_{n-}}{4S}\,,
		\end{aligned}
	\end{equation}
where $S_{n+}$ and and $S_{n-}$ are the chiral components of the spin deviation\footnote{When the Hamiltonian is quantized, the operator $S_{n+}$ creates a spin deviation at the site $n$ and the operator $S_{n-}$ destroys it}.  This leaves us with a quadratic Hamiltonian of the form
	\begin{equation} 
		H \approx \sum_{n\alpha,n'\beta} h_{n\alpha,n'\beta}S_{n\alpha}S_{n'\beta}\,,
	\end{equation}
where $\alpha$ and $\beta$  take values in $\{-,+\}$.  

The site diagonal part $H$-matrix is given by
	\begin{equation}
		\begin{aligned}
			h_{n\alpha,n \beta}=\left\{-J\frac{c_{n-1}+c_n}{4} +D\frac{2\cos^2 \theta_n-\sin^2\theta_n}{4}\right\}[\sigma_x]_{\alpha\beta}
			\\+D\frac{\sin^2\theta_n}{4}\delta_{\alpha\beta} \qquad \qquad 
			\qquad \qquad  \qquad  \qquad  \qquad \,,
		\end{aligned}
	\end{equation}
	
and the off-diagonal parts are	
	\begin{eqnarray}
		h_{n\alpha,n+1 \beta}&=&J\left(\frac{1+c_n}{4} [\sigma_x]_{\alpha\beta}+\frac{1-c_n}{4}\delta_{\alpha\beta}\right)\\
		h_{n\alpha,n-1 \beta}&=&J\left(\frac{1+c_{n-1}}{4} [\sigma_x]_{\alpha\beta}+\frac{1-c_{n-1}}{4}\delta_{\alpha\beta}\right)\,,
	\end{eqnarray}
where  $c_n\equiv \cos(\theta_n-\theta_{n+1})$ and $\sigma_i$ are the Pauli matrices. 	
	
	\subsection{\label{sec:citeref}Linearized equation of motion}

Oscillations of the spin about the equilibrium DW configuration are governed by the equation of motion
\begin{equation}
		\hbar\dot{S}_{n, \alpha}=i [H,S_{n\alpha}]\,,
\end{equation}
which results in

	\begin{eqnarray}
	-i\hbar\omega S_{n\alpha}= \qquad \qquad \qquad \qquad \qquad \qquad \qquad \qquad \qquad \qquad 
		 \\
		 i\sum_{n'\beta,n''\gamma}h_{n'\beta,n''\gamma}\left\{[S_{n'\beta},S_{n\alpha}]S_{n''\gamma}+ S_{n'\beta}[S_{n''\gamma},S_{n\alpha}]\right\}\,.\nonumber 
	\end{eqnarray}
	
	By using the  Poisson bracket (or commutator)  $[S_{n\alpha},S_{n'\beta}]=-2  S_{nZ}\epsilon_{\alpha\beta}\delta_{n,n'}$, where  $\epsilon_{\alpha\gamma}=i[\sigma_y]_{\alpha\gamma}$, and replacing $S_{nZ}$ by its equilibrium value  we obtain the linearized equation of motion for small oscillation about the equilibrium DW configuration:
	\begin{equation} \label{schero_spin_wave}
		\hbar\omega S_{n\alpha} = \sum_{n'\alpha\beta}H_{n\alpha,n'\beta} S_{n'\beta}\,.
	\end{equation}
Here the diagonal part of the spin wave Hamiltonian is expressed as:
	\begin{eqnarray}
		H_{n\alpha,n \beta}&=& -JS(c_{n-1}  +c_n)[\sigma_z]_{\alpha\beta} \nonumber \\  &+&   DS(2\cos^2 \theta_n-\sin^2\theta_n) [\sigma_z]_{\alpha\beta}\nonumber\\
		&+& DS\sin^2\theta_n[i\sigma_y]_{\alpha\beta}\,,
	\end{eqnarray}
	and the off-diagonal part is
	
	\begin{eqnarray}
		H_{n\alpha,n+1 \beta}&=&JS\left\{\frac{1+c_n}{2} [\sigma_z]_{\alpha\beta}+\frac{1-c_n}{2}[i\sigma_y]_{\alpha\beta}\right\}\,,\nonumber\\
		H_{n\alpha,n-1 \beta}&=&JS\left\{\frac{1+c_{n-1}}{2} [\sigma_z]_{\alpha\beta}+\frac{1-c_{n-1}}{2}[i\sigma_y]_{\alpha\beta}\right\}\,. \nonumber
	\end{eqnarray}

	\begin{figure}
		\centering
		\includegraphics[width=86mm]{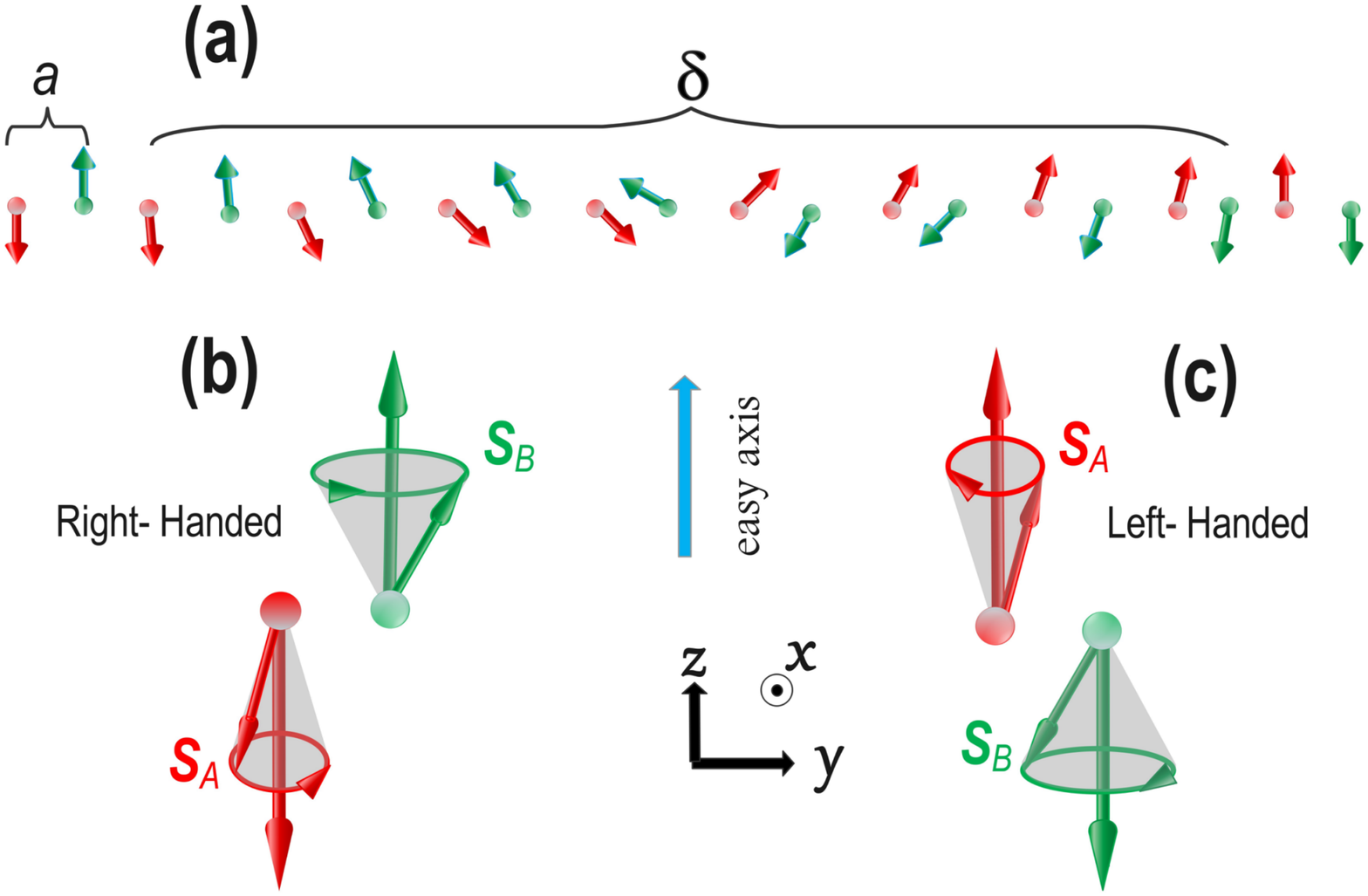}\caption{(a) Schematic of a Neel DW with a width  $\delta$. Red (green) arrows represent sublattice A (B).  Schematic of (b) RH   and (c) LH  eigenmode.   Notice that in an RH wave the amplitude of oscillations on the sublattice with up spin  is larger than on the sublattice with down spin .  This causes the RH wave
				carries a small net spin in the negative z direction.  For LH waves the situation is reversed.}	 
	\end{figure}
	
	\subsection{\label{sec:citeref}Spin waves in a uniform ground state}
	
	 Before studying spin waves on top of a domain wall, let us begin by solving the equations of motion ~(\ref{schero_spin_wave}) for a homogeneous antiferromagnetic state, described by $\theta_n=0$ for even $n$ and $\theta_n=\pi$ for odd $n$. We find two orthogonal solutions in the form of  plane waves.  The right-handed solution (written in the chiral basis) is  
	\begin{equation}
	\psi^{(RH)}_k(n) = \left\{u^{(RH)}_{k,\uparrow}\delta_{\bar n,0}+u^{(RH)}_{k,\downarrow}\delta_{\bar n,1}\right\}e^{ikna}\,,
	\end{equation}
	where  $\bar n\equiv{\rm Mod[n,2]}$ and
	\begin{equation}\label{Eigenvector1-bis}
		u^{(RH)}_{k,\uparrow}=\left(\begin{array}{c}1\\0\end{array}\right)\,,~~~~u^{(RH)}_{k,\downarrow}=-\rho_k\left(\begin{array}{c}0\\ 1\end{array}\right)\,.
	\end{equation}
Similarly, the left-handed solution  is  
	\begin{equation}
	\psi^{(LH)}_k(n) = \left\{u^{(LH)}_{k,\uparrow}\delta_{\bar n,0}+u^{(LH)}_{k,\downarrow}\delta_{\bar n,1}\right\}e^{ikna}\,,
	\end{equation}
	where
	\begin{equation}\label{Eigenvector2-bis}
		u^{(LH)}_{k,\uparrow}=\left(\begin{array}{c}0\\1\end{array}\right)\,,~~~~u^{(LH)}_{k,\downarrow}=-\rho_k^{-1}\left(\begin{array}{c}1\\0\end{array}\right)\,.
	\end{equation}
	Here   $\uparrow (\downarrow)$ refer to direction of the spins at the sublattice A(B)    and $\rho_k = \sqrt{\frac{2(J+D)-\hbar\omega_k}{2(J+D)+\hbar\omega_k}} $  is the ratio of the amplitude of the oscillation on site $A$ relative to the amplitude of the oscillations on site $B$ in the RH eigenmode.    	 
	In the absence of an external magnetic field the two eigenmodes are degenerate with eigenvalue
	\begin{equation}
\hbar\omega_k= 2 \sqrt{D(2J+D)+J^2\sin^2 ka}\,.
\end{equation} 

 In the  $RH$ mode, both up and down spins undergo a counterclockwise  precession when viewed from the  $+z$ direction with frequency $\omega_k$.  In the $LH$ mode, they both undergo a clockwise rotation with the same frequency.  A schematic illustration of two modes are shown in Fig.(1).
 
 \subsection{\label{sec:citeref}Scattering problem}
 We are now ready to formulate the scattering problem. Deep inside the region \rom{1},  $n\leq 0$, where the spins at even sites point in the $+z$ direction and the spins at odd sites point in the $-z$ direction, the  solution is taken to be of the form
	\begin{eqnarray}\label{ansatz1}
		\psi_k(n) &=& \left\{u^{(RH)}_{k,\uparrow}\delta_{\bar n,0}+u^{(RH)}_{k,\downarrow}\delta_{\bar n,1}\right\}e^{ikna}   \nonumber  \\  
		&+&r_1\left\{u^{(RH)}_{-k,\uparrow}\delta_{\bar n,0}+u^{(RH)}_{-k,\downarrow}\delta_{\bar n,1}\right\} e^{-ikna} \nonumber \\
		&+&r_2\left\{u^{(LH)}_{-k,\uparrow}\delta_{\bar n,0}+u^{(LH)}_{-k,\downarrow}\delta_{\bar n,1}\right\} e^{-ikna}\,.
	\end{eqnarray}
	This is the superposition of an incoming RH wave from the left and two reflected waves with RH and LH polarizations. $r_1$ and $r_2$ are the two reflection amplitudes.
	
In  region \rom{3}, $n>2N$, where $2N$ is the number of spins inside the DW,   the spins at even sites point in the $-z$ direction and the spins at odd sites point in the $+z$ direction.  The solution in this region is
	\begin{eqnarray}\label{ansatz2}
		\psi_k(n)&=&t_1\left\{u^{(RH)}_{k,\downarrow}\delta_{\bar n,0}+u^{(RH)}_{k,\uparrow}\delta_{\bar n,1}\right\} e^{ikna}
		\nonumber\\&+&t_2\left\{u^{(LH)}_{k,\downarrow}\delta_{\bar n,0}+u^{(LH)}_{k,\uparrow}\delta_{\bar n,1}\right\} e^{ikna}\,,
	\end{eqnarray}
i.e., a superposition of two transmitted waves of RH and LH polarization.  $t_1$ and $t_2$ are the two transmission amplitudes. 	

In the intermediate region II, defined by $0<n\leq 2N$, the solution  of Eq. (10) (with the known values of $\theta_n$)  is constructed numerically with boundary conditions imposed by the previous two equations (\ref{ansatz1}) and (\ref{ansatz2}) at $n=0$ and $n=2N+1$.    The reflection and transmission amplitudes are then obtained by requiring that Eq.~(\ref{ansatz1}) and Eq.~(\ref{ansatz2})  are satisfied at the two boundary points $n=0$ and $n=2N+1$. In view of the two-component character of the solution, this gives four linear equations from which $r_1,t_1,r_2,t_2$ can be determined.

	\begin{figure}
		\centering
		\includegraphics[width=86mm]{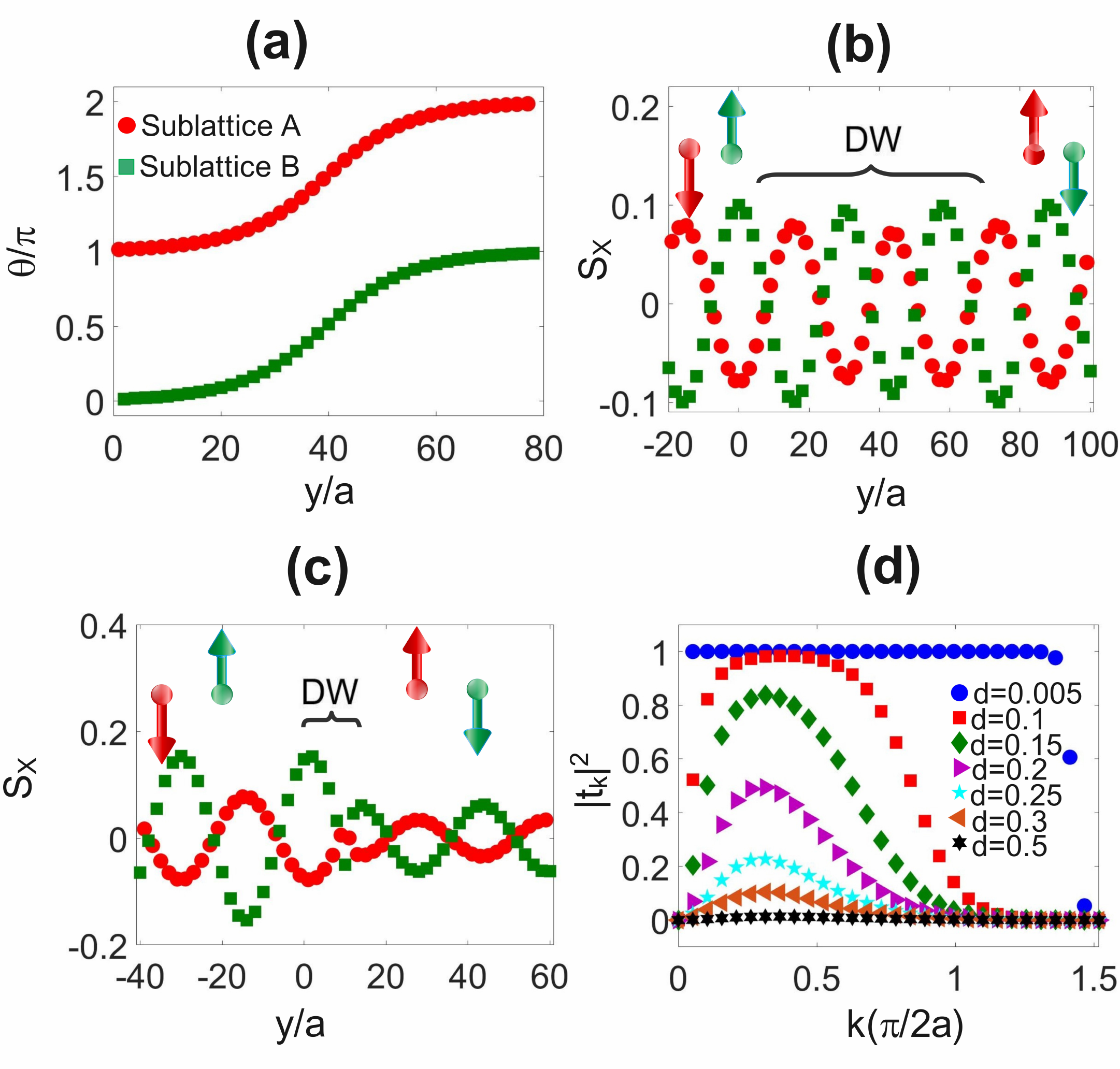}\caption{  (a) Exact DW profile
			with $d\equiv D/J=0.005$. The red dots (green squares)  represent the angles of spins at   sublattice A (B). (b) Spin wave profile for an incoming
			RH polarization with $k=0.2$ and $d=0.005$. The wave emerges with LH polarization on
			the other side of the DW. The arrows represent  the direction of the equilibrium spin at sublattice A and B on both sides of the DW. 
			(c) Spin wave profile of an incoming RH polarization  with $k=0.2$ and $d=0.2$. (d) Transmission coefficient of an incoming RH spin wave as a function of $k$ for different values of $d$. }	 
	\end{figure}

	\section{\label{sec:level1}RESULTS AND DISCUSSION}

	\subsection{\label{sec:citeref} Scattering of the spin wave from a wide DW}
	
	First, let us consider a weak anisotropy, $D/J\ll1$, in which a domain wall involves spatially slowly varying
	spins. In this limit, the N\'eel order parameter $\textbf{n}=\frac{\textbf{S}_A-\textbf{S}_B}{2}$,  is a three-dimensional unit vector and a continuous function of the
	coordinate $y$.   Assuming that, at equilibrium,  $n$ lies in the $(y,z)$ plane, we represent the equilibrium configuration of the domain wall as $\textbf{n}_{eq}=\sin(\theta){\bf y}+ \cos(\theta){\bf z}$ where the angle between ${\bf n}$ and the $z$ axis is  $\theta=2\arctan[\exp (y/\delta)]$ where $\delta=a \sqrt{J/2D}$ is the characteristic  width of the domain wall \cite{tveten2016intrinsic}. 
	
	The equation of motion for spin wave can be recast as a Schr\"odinger equation in an effective (Poschl-Teller) potential:

	\begin{equation}
		\frac{\hbar^2}{D^2}\frac{\partial ^2 \phi(y,t)}{\partial t^2}=\delta^2\frac{\partial ^2 \phi(y,t)}{\partial y^2}-\left[1-2 \sech ^2 \frac{y}{\delta}\right]\phi(y,t).
	\end{equation}
	
	Here $\phi=n_X+i n_Y$, and $n_X$,$n_Y$ are two components of the N\'eel order parameter in a {\it local} coordinate system such that the ${Z}$ axis (not to be confused with the absolute ${z}$ axis) coincides with the direction of ${\bf n_{eq}}$. The solution of this equation is 
	
	\begin{equation} 
		\phi(y,t)= \Phi\frac{\tanh \frac{y}{\delta}-i\delta k}{-1-i\delta k}e^{-i\omega t+iky}\,, ~~~ \frac{\hbar^2\omega^2}{D^2}=1+(k\delta)^2\,,
	\end{equation}
	which describes a circularly polarized wave of amplitude $\Phi$.
	The profile of a spin wave on top of a wide DW is shown in Fig. 2(b). We see that an incoming RH spin wave  passes through the DW without reflection and emerges  on the other side with LH polarization \cite{kim2014propulsion}.  We  also notice that, in the left hand side of the DW, the amplitude of  the oscillation on sublattice with up spin (green squre) is larger than that of sublattice with down spin (red dots). The situation becomes reversed in the right hand side of the DW, indicating a change of the polarization of transmitted spin wave.   Completely analogous results are obtained,  , for an incoming LH polarization.

	As $D/J$ increases the DW becomes progressively narrower and the continuum approximation breaks down. The numerical spin wave solutions of an incoming RH spin wave for $D/J=0.2$ and $k=0.2$ are shown in Fig. 2(c). The RH spin wave  is partially reflected without change in polarization. However, the transmitted spin wave absorbs  spin angular momentum from the DW and reverses its polarization upon transmission  \cite{kim2014propulsion}.   The scenario for an incoming LH spin wave is  similar. In this case the LH spin wave transfers spin angular momentum to the DW and emerges as a RH spin wave on the other side of the DW.  
	
	The $k$-dependent transmission coefficient of an incoming RH spin wave are shown in Fig. 2(d), for different values of $D/J$.  When $D/J\ll1$ the transmission coefficient is $|t_k|^2\approx 1$ for a wide range of $k$. For larger $D/J$ the wave is partially reflected, and the transmission coefficients has a Gaussian-type shape  with full width at half maximum in the range of $0.2<k<0.5$. This suggests that magnons with relatively large wave vector will dominate the spin transport  in an uniaxial AF  with  large anisotropy.

	\subsection{\label{sec:citeref}  Scattering of the Spin Wave from a Sharp DW	 }
	
	Starting at $D/J=2/3$ and for all larger values of $D/J$   the equilibrium configuration of the DW becomes abruptly  sharp and the scattering pattern of spin waves becomes different\cite{barbara1973proprietes,barbara1994magnetization}.  See Fig.3 for the spin configurations of an abrupt domain wall. First of all, we notice that an abrupt domain wall, unlike a smooth one, has a net spin associated with it: we can regard it as a small ferromagnetic insertion in an antiferromagnetic background, and the spin of this insertion can point up or down.  Second, the DW configuration is now invariant under rotations about the $z$ axis, implying that  different polarizations do not mix: an RH wave cannot be converted to an LH wave and vice versa; reflection and transmission amplitudes are strictly diagonal in the polarization index.  We will see that the transmission coefficient depends strongly on the polarization of the incoming wave relative to the orientation of  spins at the domain boundary.
		\begin{figure}
		\centering
		\includegraphics[width=86mm]{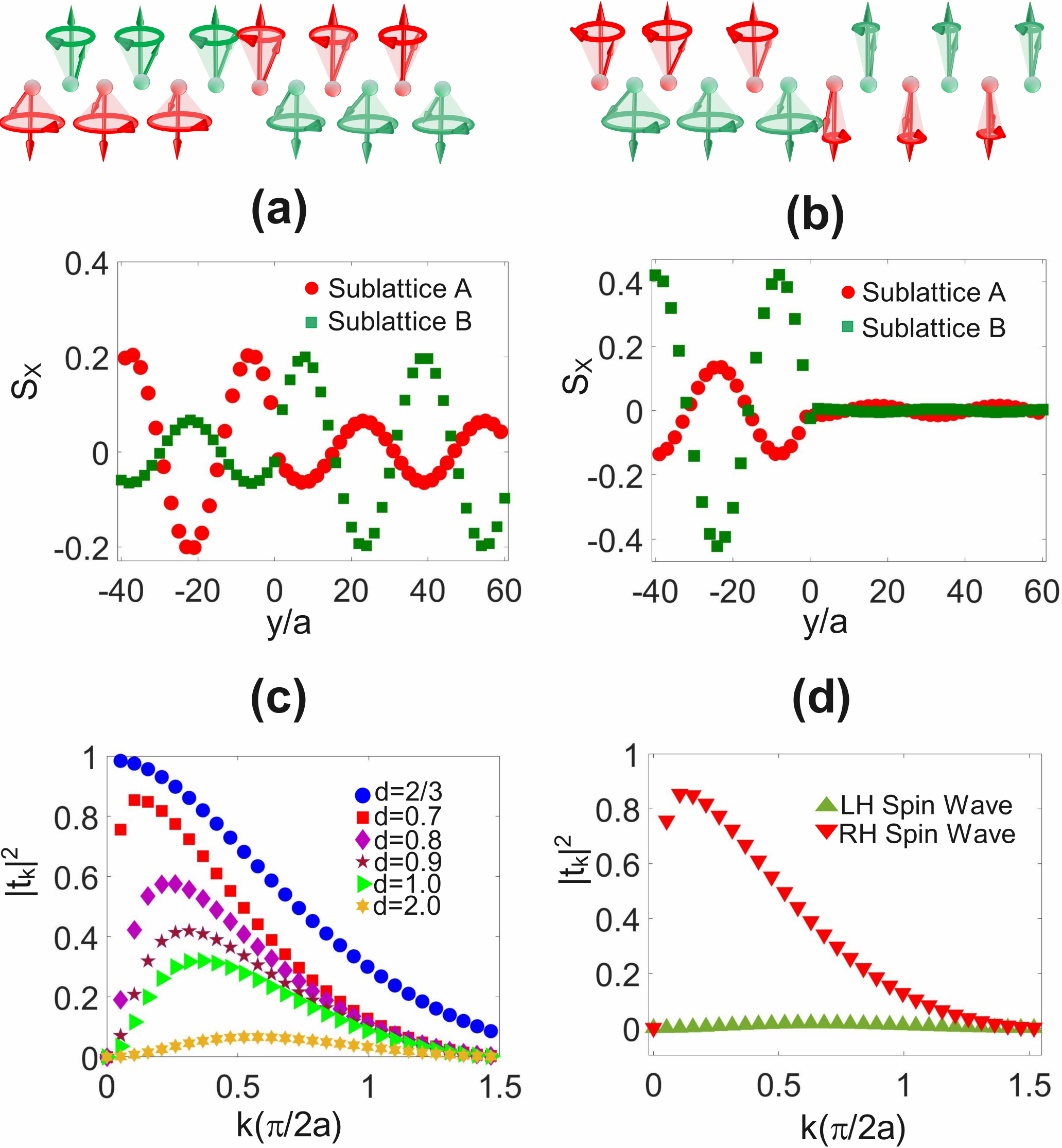}\caption{ (a) Spin wave profile of an incoming LH polarization which scatters from an abruptly sharp DW with two up-spins at the domain boundary at $d=0.67$.   The wave is transmitted with small reflection.   (b)  Same as in   (a)  for an abruptly sharp DW with two down-spins at the domain boundary. Only a small portion of the wave is  transmitted. (c) Transmission coefficient  of an incoming LH spin waves scattered by two up-spin at the domain boundaries for different values of $D/J\equiv d$. (d) Transmission coefficient of an RH spin wave (red triangles) and an LH spin wave (green triangles) for an abruptly sharp DW with two down-spins at the domain boundary as a function of $k$ at $d =0.7$.  }	 
	\end{figure}	
	
	The appeal of the sharp domain wall configuration is that it admits a completely analytical solution for the reflection and transmission amplitudes of spin waves.  By substituting the Ansätze from Eq.(\ref{ansatz1}) and Eq.(\ref{ansatz2}) into Eq.(\ref{schero_spin_wave})  for an abrupt DW with two up spins the reflection and transmission amplitudes of an RH wave work out to be
	\begin{equation}\label{RRH-sharp}
		r_k^{RH}=\frac{-1+| \rho_k e^{-i ka} -2 D/J+\hbar\omega_k/J |^2}{1-\left(\rho_k e^{i ka} -2 D/J+\hbar\omega_k/J \right)^2}
	\end{equation}
	and
	\begin{equation}\label{TRH-sharp}
		t_k^{RH}=  \left(r_k^{RH}+e^{-2 i ka}\right)\rho_k-e^{-i ka} (2 \dfrac{D}{J}-\dfrac{\hbar\omega_k}{J} ) (r_k^{RH}+1)
	\end{equation}
	where $\omega_k$ is the frequency of the incoming wave.
	For the same DW with two up spins the reflection and transmission amplitudes of a LH wave are obtained by simply replacing $\omega_k \to -\omega_k$ in the above formulas, i.e., explicitly
	\begin{equation}\label{RLH-sharp}
		r_k^{LH}=\frac{-1+| \rho_k^{-1} e^{-i ka} -2 D/J-\hbar\omega_k/J |^2}{1-\left(\rho_k^{-1} e^{i ka} -2 D/J-\hbar\omega_k/J \right)^2}
	\end{equation}
	and
	\begin{equation}\label{TLH-sharp}
		t_k^{LH}=\left(r_k^{LH}+e^{-2 i ka}\right)\rho^{-1}_k-e^{-i ka} (2 \dfrac{D}{J}+\dfrac{\hbar\omega_k}{J} ) (r_k^{LH}+1)
	\end{equation}
	The reflection and transmission amplitudes for a sharp DW with two down spins at the boundary 
	can be obtained from the above results by interchanging the right-hand and the left-hand polarizations. The analytic expressions~(\ref{RRH-sharp}-\ref{TLH-sharp})  for the reflection  and  transmission amplitudes of spin waves of two polarizations interacting with an abrupt antiferromagnetic DW are one of the main results of this paper.
		\begin{figure}
		\centering
		\includegraphics[width=86mm]{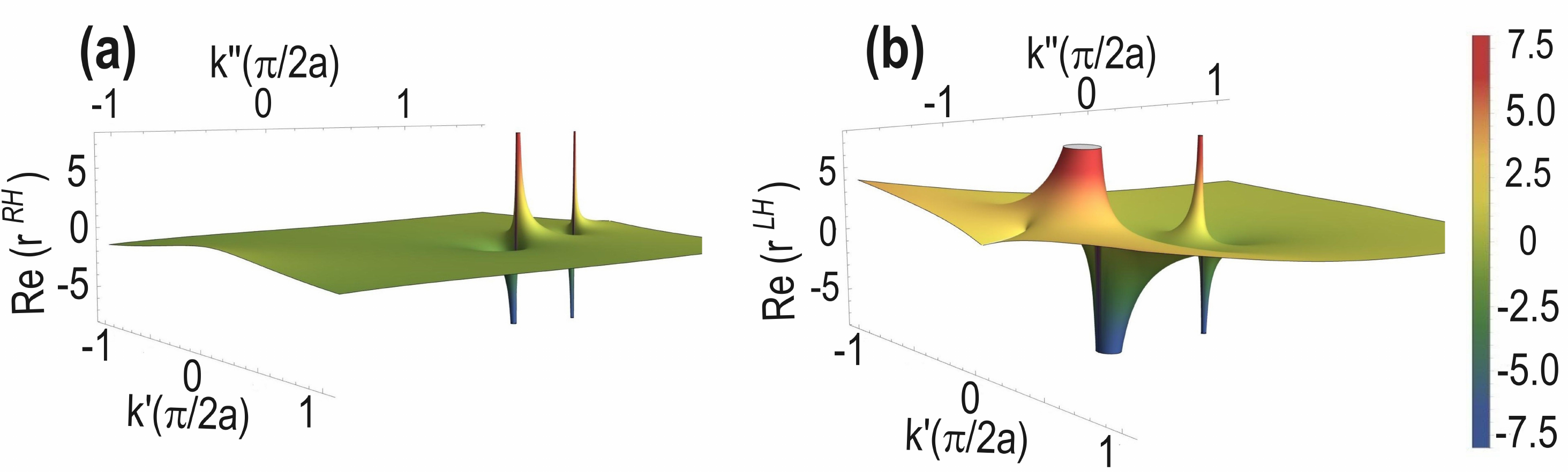}\caption{(a),(b) Real part of transmission amplitude of an incoming RH(LH) spin wave in a complex plane, i.e. $k=k^\prime+i k^{\prime\prime}$ for $D/J=1$.}	 
	\end{figure}
	
	Armed with these analytical results we can easily plot the profile of spin waves as they get scattered by an abrupt DW.
Let us first consider an incoming LH spin wave. As we can see in Fig. 3(a) for $k=0.2$ and $D/J=0.67$ the spin wave is  transmitted through the DW without change in polarization. As $D/J$ becomes larger the transmission coefficient becomes smaller until, at $D/J\approx4$, it becomes almost zero for the entire range of $k$ (see Fig. 3(c)). In contrast, for a sharp DW with two down-spin at the domain boundary, the LH spin wave  is mostly    reflected for all values of $k$.(Fig 3. (d)) 
	
	In the case of  an incoming RH spin wave, the wave can be partially transmitted to the other side of a DW with two down-spin at the domain boundary, but is  mostly reflected by a two up-spin DW.  We see that changing the polarization of the incoming spin wave while simultaneously flipping the spins in the DW is a symmetry of the system.

	\begin{figure}
		\centering
		\includegraphics[width=80mm]{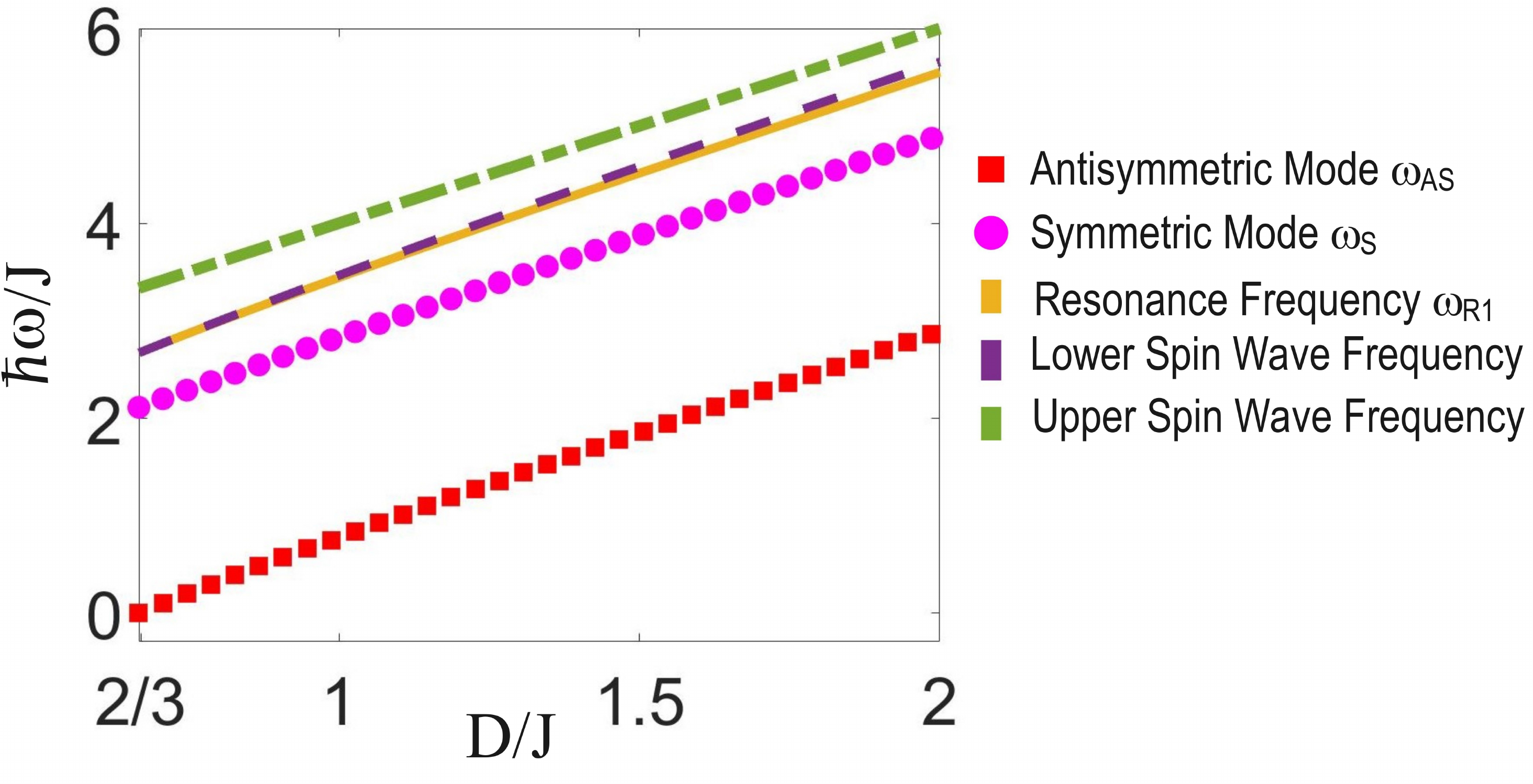}\caption{Spin wave frequency of a sharp DW as a function of $D/J$. Red squares  show antisymmetric  bound states and purple circles show symmetric bound states. The yellow solid line shows the  resonance frequency ($\omega_{R1}$). The purple dashed line represents the lower bound of the continuous spin wave frequency range, while the green dashed line represents the upper bound of that range.}	 
	\end{figure}

	We gain some insight into this surprising polarization dependence of the transmission coefficient by considering the analytic structure of the transmission amplitude for complex $k$ (See Fig. 4). In the case of an up-spin DW the transmission amplitude for RH waves has two poles on the positive imaginary axis $k\rightarrow i\kappa$ with $\kappa>0$, which correspond to bound states , i.e., spin oscillations that are strongly localized on top the sharp DW and decay exponentially away from the DW.  These bound states are RH polarized and their frequencies -- obtained by substituting $k\rightarrow i\kappa$ in the formula for the spin wave frequency -- are given by
	\begin{equation}
		\hbar\omega_S= 2\sqrt{D(J+D)}	
	\end{equation}
	and,
	\begin{equation}
		\hbar\omega_{AS}= \frac{-4J+\sqrt{36D^2+12DJ-8J^2}}{3}	 
	\end{equation}
	where the subscript S refers to symmetric oscillation of the spins (i.e., the two spins at the center of the DW rotate in phase) while the subscript AS refers to antisymmetric oscillations of the spins (i.e., the two spins at the center of the DW rotate with a $\pi$ phase difference between them). Notice that the antisymmetric mode frequency vanishes at $D/J=2/3$.

	
	
	The transmission amplitude for LH waves has no simple poles on the positive imaginary $k$-axis, but it does have two poles on the negative imaginary $k$-axis, i.e., at  $k=i\kappa$, with $\kappa<0$.   The associated frequencies of these two poles are:
	
	\begin{equation}
		\hbar\omega_{R1}=\frac{4J+\sqrt{36D^2+12DJ-8J^2}}{3}	~~,~~  \hbar\omega_{R2}=   \hbar\omega_S	\,.   
	\end{equation}
	
	What is the physical meaning of these two poles on the negative imaginary $k$-axis?  They are not bound states, because the associated solution of the equation of motion would diverge exponentially away from the DW. Nevertheless, the existence of such poles can have a large impact on the transmission coefficient if the pole occurs sufficiently near to the real axis \cite{nakayama2019perfect}. When this happens, the pole is described as a {\it resonance}, meaning that it can cause a rapid increase of the transmission coefficient in a relatively narrow range of real wave vectors close to the pole.   How narrow this range is  depends on how close the pole is to the real axis.  In the present case, the pole at $\omega_{R1}$ is associated with an imaginary wave vector that occurs extremely near to the real axis when $D/J$ approaches $2/3$. This is the mathematical origin of the sharp peak in the transmission coefficient of LH waves for small real $k$ and for $D/J$ close to 2/3. For larger real $k$ and/or for $D/J \gg 2/3$ the resonance disappears, as we clearly see in the numerical results (Fig 3.(c)).
	
	A more physical description of the phenomenon can be achieved by noting that for $D/J \rightarrow 2/3$ the resonant frequency $\omega_{R1}$ approaches  the real spin wave frequency $\omega_{k=0}$ (See Fig. 5).  We can therefore say that for $D/J\approx 2/3$ and $k\approx0$ the frequency of the incoming spin wave matches the frequency of the LH resonant state, resulting in a peak of the transmission coefficient.
	
	To corroborate this point of view we have derived an expression for the transmission probability of a spin wave in terms of bound states and resonance frequencies:
	\begin{equation}\label{TKFaridi}
		|t_k|^2=\frac{J^2((2 D+2J)^2-\hbar^2\omega_k^2 )  \left(\hbar^2\omega_k ^2-4 D (D+2J)\right)}{3\hbar^4 (\pm\omega_k^2 -\omega_S^2) (\pm\omega_k -\omega_{AS})  (\pm\omega_k +\omega_{R1})}\,,
	\end{equation}
	where the upper sign applies to RH waves and the lower sign to LH waves.  We emphasize that this expression is positive with values in the range $0<|t_k|^2<1$ for all real $k$.
	
	
	Now we can understand the selective transmission of spin waves according to their polarizations. In the case of incoming $LH$ spin waves, and for $D/J\approx 2/3$, the spin wave frequency $\omega_k$, stays close to the resonance frequency i.e., for $\omega_k\approx \omega_{R1}$.  Therefore, the transmission probability becomes large (however, the vanishing numerator of Eq.~(\ref{TKFaridi})  prevents divergence when $\omega_k=\omega_{R1}$). As $D/J$ increases the difference between $\omega_k$ and $\omega_{R1}$ becomes larger, which results in a decreasing  transmission probability. In contrast to LH, an RH spin wave  of frequency $\omega_k$ never gets close to resonance due to the different sign in the denominator of Eq.~(\ref{TKFaridi}): this explains the negligible transmission probability of RH spin waves.

	\section{\label{sec:level1}SUMMARY AND OUTLOOK}

	  In conclusion, we have theoretically demonstrated that a sharp antiferromagnetic DW can act as a filter for the polarization of spin waves. As the DW becomes abruptly sharp, the state of circular polarization of an incoming wave (RH or LH) remains unchanged in the transmission/reflection process.  For suitable values of the anisotropy parameter the DW allows  one of the two polarizations to pass while largely reflecting the other.  A RH-polarized  incoming spin wave gets mostly reflected by an abruptly sharp DW with two up-spins at the center, but it can be  partially transmitted through a DW with two down-spins.  Conversely,  a LH-polarized  incoming spin wave gets totally reflected by an abruptly sharp DW with two down-spins  at the center, but it can be  partially transmitted through a DW with two up-spins.  We understand these results in terms of resonant states (i.e., poles of the transmission amplitude for $k$ close to the real axis but on the {\it negative} imaginary axis) whose frequency almost   matches the frequency of the incoming wave. These resonances occur for one polarization but not for the other.  With an eye towards applications, these findings suggest that  atomically sharp domain walls can be used in magnonic circuits as  spin polarizers.
	\begin{acknowledgments}
	S.K.K. was supported by Brain Pool Plus Program through the National Research Foundation of Korea funded by the Ministry of Science and ICT (NRF-2020H1D3A2A03099291), by the National Research Foundation of Korea(NRF) grant funded by the Korea government(MSIT) (NRF-2021R1C1C1006273), and by the National Research Foundation of Korea funded by the Korea Government via the SRC Center for Quantum Coherence in Condensed Matter (NRF-2016R1A5A1008184).
	\end{acknowledgments}
	
	\appendix
	\section{ANTIFERROMAGNETIC SPIN WAVE-THE HOMOGENEOUS SOLUTION}

	In this appendix  we find the eigenfunctions and eigenvalues of a homogeneous antiferromagnetic structure. To do so, we set $\theta_n=\pi \bar n$,  where $\bar n\equiv{\rm Mod[n,2]}$.  This is $0$ on the even sites and $\pi$ on the odd sites.   Then the spin wave Hamiltonian takes the form:
	\begin{eqnarray}
		H_{n\alpha,n \beta}&=&\left\{2J+2D\right\}S[\sigma_z]_{\alpha\beta} \nonumber \\
		H_{n\alpha,n+1 \beta}&=&JS[i\sigma_y]_{\alpha\beta} \nonumber \\
		H_{n\alpha,n-1 \beta}&=&JS[i\sigma_y]_{\alpha\beta}\,.
	\end{eqnarray}
The solution has the form
	\begin{equation}
		\psi_{k}(n)=u_k(n)e^{ikna}\,,
	\end{equation}
where $u_k(n)$ is a two-component spinor which satisfies the periodicity condition $u_k(n+2)=u_k(n)$ and $k$ is in the range $-\frac{\pi}{2a}<k<\frac{\pi}{2a}$  .  This means that $u_k(n)$ has only two distinct values, which we denote by $u_{k,0}\equiv u_k(0)$ and $u_{k,1}\equiv u_k(1)$.  This function can be written as
	\begin{equation}
		u_k(n)=u_{k,0}\delta_{\bar n,0}+u_{k,1}\delta_{\bar n,1}\,.
	\end{equation}
Notice that, with these definitions, we have
	\begin{equation}
		\psi_{k}(0)= u_{k,0}\,,~~~~\psi_k(1)=u_{k,1}e^{ika}\,.
	\end{equation}
$u_{k,0}$ and $u_{k,1}$ are determined by applying the equation of motion to the sites $n=0$ and $n=1$ in the unit cell. Thus we get
	\begin{equation}
		\hbar\omega_k u_{k,0}= 2(J+D) \sigma_z \cdot u_{k,0} + 2J\cos (ka)(i\sigma_y)\cdot u_{k,1}
	\end{equation}
	and,
	\begin{equation}
		\hbar\omega_k  u_{k,1}= 2(J+D) \sigma_z\cdot u_{k,1} +2J\cos(ka)(i\sigma_y)\cdot u_{k,0}\,.  
	\end{equation}

We have two doubly degenerate eigenvalues
	\begin{equation}
		\hbar\omega_k=\pm2 \sqrt{D(2J+D)+J^2\sin^2 ka}\,.
	\end{equation} 
	
A possible choice of  degenerate eigenvectors (for positive frequency) is
	\begin{equation}\label{Eigenvector1-bis}
		u^{(RH)}_{k,0}\equiv u^{(RH)}_{k,\uparrow}= \left(\begin{array}{c}1\\0\end{array}\right)\,,~~~~u^{(RH)}_{k,1}\equiv u^{(RH)}_{k,\downarrow}=-\rho_k\left(\begin{array}{c}0\\ 1\end{array}\right)
	\end{equation}
	and
	\begin{equation}\label{Eigenvector2-bis}
		u^{(LH)}_{k,0} \equiv u^{(LH)}_{k,\uparrow}=\left(\begin{array}{c}0\\1\end{array}\right)\,,~~~~u^{(LH)}_{k,1}\equiv u^{(LH)}_{k,\downarrow}=-\rho_k^{-1}\left(\begin{array}{c}1\\0\end{array}\right)\ 
	\end{equation}.

\section{Spin Wave Solution for Inhomogeneous Spin Chain}
	Here we illustrate the numerical method for solving the equation of motion in the inhomogeneous region (DW).
	
	The DW  includes the sites $n=1,..., 2N$, where the values of $\theta_n$ are already determined from Eq.(\ref{DW_profile}). We first show that the amplitudes $\psi_1,...,\psi_{2N}$ can be expressed as linear functions of $\psi_0$ and $\psi_{2N+1}$.  
	
	 To this end define the propagator
	\begin{equation}
	G=(\hbar\omega \tilde 1 - \tilde H)^{-1}\,,
	\end{equation} 

	where $\tilde 1$ is a $4N \times 4N$ identity matrix and $\tilde H$ (also a $4N\times 4N$ matrix) is the restriction of the Hamiltonian to the subspace of the transition sites $1,...,2N$.  Then, for $1\leq n\leq 2N$ we have
	
	\begin{equation}
		\psi_n = G_{n,2N}H_{2N,2N+1}\psi_{2N+1}+G_{n,1}H_{1,0}\psi_0\,,~~~1\leq n \leq 2N\,.
	\end{equation}

	In addition the solutions for $\psi_n$ in the $n\leq 0$ and $n\geq 2N+1$ are given by Eq. (\ref{ansatz1}) and Eq. (\ref{ansatz2}) respectively. 
	
	These formulas guarantee that the equation of motion is satisfied identically (i.e., for any choice of $r_1,t_1,r_2,t_2$) at almost every site, with the exception of the two sites $n=0$ and $n=2N+1$.  On these special ``frontier" sites the equation of motion is satisfied only for a specific choice of $r_1,t_1,r_2,t_2$.  Thus, the equations that determine the four scattering amplitudes are
	
\begin{eqnarray}
(\hbar\omega-H_{0,0})\psi_{0}-H_{0,-1}\psi_{-1}-H_{0,1}\psi_{1}=0\nonumber\\
(\hbar\omega-H_{2N+1,2N+1})\psi_{2N+1}-H_{2N+1,2N}\psi_{2N}-\\ \nonumber H_{2N+1,2N+2}\psi_{2N+2}=0.
\end{eqnarray} 
	
	Let us insert the formulas for $\psi_1$ and $\psi_{2N}$:

	\begin{eqnarray}
	\psi_1 &=&G_{1,2N}H_{2N,2N+1}\psi_{2N+1}+G_{1,1}H_{1,0}\psi_0, \nonumber\\
	\psi_{2N} &=&G_{2N,2N}H_{2N,2N+1}\psi_{2N+1}+G_{2N,1}H_{1,0}\psi_0\,.
	\end{eqnarray}
	This gives us the equations
	\begin{eqnarray}
(\hbar\omega-H_{0,0}-H_{0,1}G_{1,1}H_{1,0})\psi_0 \nonumber -H_{0,-1}\psi_{-1}&-&\\ \nonumber H_{0,1}G_{1,n}H_{n,n+1}\psi_{n+1}=0 \nonumber
\end{eqnarray}
and
\begin{eqnarray}
\nonumber (\hbar\omega-H_{2N+1,2N+1}&-&\\ \nonumber H_{2N+1,2N}G_{2N,2N}H_{2N,2N+1})\psi_{2N+1}&-&\\ \nonumber H_{2N+1,n}G_{2N,1}H_{1,0}\psi_0-H_{2N+1,2N+2}\psi_{2N+2}&=&0\,.\nonumber\\ 
	\end{eqnarray}
All the quantities that appear in these equations are expressed in terms of $r_1,t_1,r_2,t_2$ and there are four equation because $\psi$ is a two-component spinor. These equations can be solved to yield the scattering amplitudes.

	\section{Bound States}
	
	In order to obtain an expression for each of the two bound states frequency, first we introduce the Ansätze for  $n\leq0$:
	\begin{eqnarray} \label{e-wave}
		\psi_n(\kappa)= r_1\left\{u^{(RH)}_{k,\uparrow}\delta_{\bar n,0}+u^{(RH)}_{k,\downarrow}\delta_{\bar n,1}\right\} e^{\kappa na} \nonumber\\
		+r_2\left\{u^{(LH)}_{\kappa,\uparrow}\delta_{\bar n,0}+u^{(LH)}_{\kappa,\downarrow}\delta_{\bar n,1}\right\} e^{\kappa na},
	\end{eqnarray}
	and for $ n\geq1$
	\begin{eqnarray}
		\psi_n(\kappa)=t_1\left\{u^{(RH)}_{\kappa,\downarrow}\delta_{\bar n,0}+u^{(RH)}_{\kappa,\uparrow}\delta_{\bar n,1}\right\} e^{-\kappa na}\nonumber \\+t_2\left\{u^{(LH)}_{\kappa,\downarrow}\delta_{\bar n,0}+u^{(LH)}_{\kappa,\uparrow}\delta_{\bar n,1}\right\} e^{-\kappa na}\,. \nonumber
	\end{eqnarray}
These are  obtained by replacing  $k\rightarrow i\kappa$ in  Eq.(\ref{ansatz1}). 

	By substituting Eq.(\ref{e-wave}), as well as the dispersion relation of an evanescent wave,  $\hbar\omega_\kappa=2 \sqrt{D(2J+D)-J^2\sinh^2 \kappa a}$, in Eq.(\ref{schero_spin_wave}) with $n=0$ we obtain
	\begin{equation} \label{EOMpsi0}
		H_{0\alpha,0\beta} \psi_0+H_{0\alpha,-1\beta} \psi_{-1}	+H_{0\alpha,1\beta} \psi_1=\hbar\omega \psi_0\,.  
	\end{equation}	 
	
	The two spins at the domain boundaries can oscillate symmetrically or antisymmetrically relative to each other.  In the symmetric mode, the two spins oscillate with the same amplitude and with the same  phase ($\psi_0=\psi_1$),  while in the antisymmetric mode they oscillate with the same amplitude but with opposite phase ($\psi_0=-\psi_1$). Then Eq.(\ref{EOMpsi0}) can be simplified to: 
	\begin{equation} \label{EOMsimple}
		2\frac{D}{J}\sigma_z\psi_0+i\sigma_y \psi_{-1}\pm	\sigma_z \psi_0= \frac{\hbar\omega}{J}\psi_0, \qquad  n=0\,,
	\end{equation}			 
where $(+)$ and $(-)$ refer to symmetric and anti-symmetric modes respectively. 
	Finally, the solution of Eq.(\ref{EOMsimple}) gives the following expression for the frequencies of the bound states:
	\begin{equation}
		\hbar\omega_{AS}= \frac{-4J+\sqrt{36D^2+12DJ-8J^2}}{3}	 
	\end{equation}
and	
	\begin{equation}
		\hbar\omega_S= 2\sqrt{D(J+D)}\,,	
	\end{equation}
which agree with the numerical results shown in Fig.(5).
	
	\nocite{*}
	
	\bibliography{./myreferences}
	
\end{document}